\documentstyle{article}   

\title{Generating new perfect-fluid solutions from known ones: II} 
\author{Istv\'an R\'acz\thanks{e-mail: istvan@rmkthe.rmki.kfki.hu} 
{\ and}  J\'ozsef Zsigrai\thanks{e-mail: zsigrai@rmkthe.rmki.kfki.hu}
\\ {\small \it MTA KFKI Research Institute for Particle and Nuclear Physics}
\\ {\small \it H-1525 Budapest 114, P.O.B. 49, Hungary}}  

\begin{document}         

\maketitle                

\begin{abstract}

The properties
of a transformation  previously considered for
generating new perfect-fluid solutions from known ones are 
further investigated. It is assumed that the four-velocity
of the fluid is parallel to the stationary Killing field, and
also that the norm and the twist potential of the stationary
Killing field are functionally related.
This case is complementary to the case studied in our
previous paper.
The transformation can be applied to generate possibly new
perfect-fluid solutions from known ones only for the case of 
barotropic equation of state $\rho+3p=0$ or, alternatively, 
for the case of a static spacetime. For static spacetimes our method
recovers the
Buchdahl transformation. It is
demonstrated, moreover, that Herlt's technique for constructing
stationary perfect-fluid solutions from static ones is, actually, a
special case of the method considered in the present paper.
\vskip 0.5truecm
{\noindent
PACS numbers 0420J, 0440N}

\end{abstract}

\begin{section}{Introduction}
A method has been developed in  \cite{ESI,R,RZS} for
generating stationary perfect-fluid solutions from known ones. This method was introduced as a generalization of the
Geroch transformation \cite{Geroch} (see also \cite{Cosgrove}) given for
vacuum spacetimes.
 It was assumed there that the four-velocity of the fluid is
parallel to the stationary Killing field, and also, that the norm and the
twist potential of the stationary Killing field are functionally
{\it independent}. With these assumptions it was shown in \cite{RZS} that the
generalized transformation can be applied to create possibly new
perfect-fluid solutions from known ones only for the case of 
the barotropic
equation of state $\rho+3P=0$. In the present work we study the
complementary problem, that is, here we assume that the norm and the
twist potential of the stationary Killing field are functionally
{\it related}
and investigate the applicability of the generalized
transformation. It turned out that either the equation
of state of the fluid has to be the above restricted one for both the
initial and the resulted configurations, or the
spacetime has to be static. In the latter case our transformation 
reduces
to the Buchdahl transformation of perfect-fluid
spacetimes \cite{Buchdahl, Stewart}
\footnote{We have just learnt that the transformation
described in \cite{ESI,R,RZS} is a re-discoveration of a general
Lie-B\"acklund transformation studied earlier by Stephani \cite{Stephani}.
Nevertheless, due to the different approach we used to get our
transformation, there are particular results (e.g. the relation of the
generalized Geroch transformation to the Buchdahl transformation)
described in the present paper which were out of the scope of \cite{Stephani}
and deserve further attention.}.

We would like to emphasize that most of the analysis given
 in section 2 of the present paper is valid also in the functionally
independent case of \cite{RZS}. In \cite{RZS} we recalled that the functional
independence of the twist potential and the norm of the Killing field ensures that any
solution to the reduced set of equations (2.7)-(2.9) of \cite{RZS} satisfies the
complete system of  Einstein's equations. Using this fact it was possible to generalize the
Geroch transformation to the selected perfect-fluid configurations. In the present
paper, however, we use the full set of Einstein's equations in order to introduce the
generalized transformation and to study the restrictions on its applicability.
This analysis is, in turn, actually independent of any assumptions about the
functional dependence of the twist potential and norm of the
Killing field.

This paper is organized as follows. We recall the field equations for the
selected perfect fluid configurations and study their symmetry
properties. Then, we determine the restrictions on the
applicability of the generalized transformation for generating new
perfect-fluid solutions for the case under consideration. Next, we examine
the particular case of a static spacetime and determine the
transformation of the equation of state. Finally, 
we study the relationship between Herlt's technique \cite{Herlt} developed for
generating stationary perfect-fluid solutions from static ones and the method
considered in this paper.
It is shown that the method introduced by Herlt can be considered as a special case of the
one presented in this paper. In particular, it follows that the
equation of
state for both the initial and resulted perfect-fluid  configurations 
has to be $\rho+3P=0$ whenever Herlt's transformation is applied in order
to get stationary solutions from static ones
\footnote{Note, however, that  our conclusion applies merely to one of the
methods introduced by Herlt in \cite{Herlt}.}.

\end{section}

\begin{section}{Symmetry properties of the field equations}
Start with a stationary perfect-fluid
spacetime with four-velocity, $u^a$, parallel to the timelike
Killing field $\xi^a$. Then 
\begin{equation} u^a=(-v)^{1/2}\xi^a, \label{ua}\end{equation}
where $v=\xi_a\xi^a$ is the norm of the timelike Killing field.
Consequently, the flow of the fluid is ``rigid" in the sense that
it is shear- and expansion-free.
It was shown in \cite{GerochLindblom} that these perfect-fluid spacetimes
can be considered as faithful representations of equilibrium states of real, 
dissipative relativistic fluids.

The existence of a Killing field allows one to use the three-dimensional formulation
(projection formalism) of general relativity
\cite{Geroch,Exact}. Moreover,
for the perfect-fluid configurations under consideration
the twist $\omega_a$ $(:=\epsilon_{abcd}\xi^b\nabla^c\xi^d)$ of the Killing field can
always  be expressed as a gradient of a function, $\omega$, called the twist potential
of the Killing field. The field equations for stationary perfect-fluid spacetimes
having the 4-velocity of the fluid  parallel to the timelike
Killing field read

\begin{equation}H_{ab}:=R_{ab}-16\pi v^{-1}Ph_{ab}={1\over
2}v^{-2}\bigl[ (D_av)(D_bv)+(D_a\omega )(D_b\omega )\bigr]\ ,
\label{fe1} \end{equation}
\begin{equation}D_mD^mv=v^{-1}\bigl[(D_mv)(D^mv)-(D_m\omega )(D^m\omega )\bigr]-8\pi
(\rho+3P)\ ,\label{fe2} \end{equation}
\begin{equation}D_mD^m\omega=2v^{-1}(D_m\omega )(D^mv)\ , \label{fe3} \end{equation}
where $\rho$ is the mass-density and $P$ is the pressure of the fluid, while 
$R_{ab}$ and $D_a$ are the Ricci-tensor and covariant derivative associated with the
three-dimensional Riemannian metric $h_{ab}:=-vg_{ab}+\xi_a\xi_b$ (here $g_{ab}$ denotes
 the spacetime metric).
 In addition to the above equations the Euler-Lagrange equation 
\begin{equation} \partial _aP+{1\over 2} (\rho+P){{\partial _av}\over v}=0\ ,
\label{EL}\end{equation}
which  follows from the above ones, has to be satisfied.

In \cite{RZS} it was shown that whenever the twist and the gradient of the norm of
the Killing field are linearly independent then one can choose geometrically
preferred local coordinates in which the functions $v$ and $\omega$ depend merely
on two coordinates, say $x^1$ and $x^2$. In such coordinates it is straightforward to
realize the 
analogy between the field equation (\ref{fe1}) and the corresponding vacuum 
equation (for the study of the vacuum case see \cite{Geroch} and also
\cite{Cosgrove}). Both equations have a symmetry property, which allows one to
introduce a transformation for generating new solutions of Einstein's equations
from known ones. In this way, it was possible to generalize the vacuum
transformation given by Geroch \cite{Geroch} to the selected perfect-fluid spacetimes.
However, in the perfect-fluid case there are additional field equations which have
to be satisfied when applying the generalized transformation. This, in turn, 
means that the generalized transformation  of \cite{ESI,R, RZS}
 can be applied only if 
the equation of
state of the fluid is $\rho+3P=0$, for both the initial and the final
configurations.

In the present paper we consider the complementary case, when the twist,
 $\omega_a=D_a\omega$, and the gradient of the
norm, $D_av$, of the Killing field are linearly dependent.
This is possible whenever at least one of the functions $v$ 
and $\omega$ is constant throughout or there exists
a function $\omega=\omega(v)$ (or, alternatively, $v=v(\omega)$).
Provided that at least one of the functions $v$ and $\omega$ is
non-constant\footnote{If
 both $v$ and $\omega$ are constant, then we have from the field
equations that the equation of state is $\rho+3P=0$, where both $\rho$ and
$P$ are constant. Clearly, the transformation introduced bellow (see equations
(\ref{tauprime})-(\ref{omega})) yields only  a gauge transformation of these
spacetimes.}, there exists
 at any point $p$ of the three-dimensional space of Killing orbits a 
two-dimensional subspace of the tangent space spanned by geometrically preferred vectors
satisfying all the equations (3.6)-(3.9) of \cite{ESI}. These
geometrically preferred vectors are 
perpendicular to the one-dimensional subspace of the tangent space at $p$ spanned
by $D_av$ or $D_a\omega$.
 Then, in all the above cases 
one can choose, at least in an appropriately small neighborhood of $p$,
two linearly independent preferred non-vanishing
vector fields, $k^a_{(2)}$ and $k^a_{(3)}$, and introduce a local 
coordinate system $(x^1,\ x^2,\ x^3)$ adapted to these vector fields,
 so that $k^a_{(2)}=\bigl({\partial\over{\partial x^2}}\bigr)^a$ and
 $k^a_{(3)}=\bigl({\partial\over{\partial x^3}}\bigr)^a$.
Then it follows immediately that
\begin{equation} v=v(x^1) \ \ ,\ \ 
\omega=\omega(x^1)\ .\label{vomega}\end{equation}
Furthermore, in such a local coordinate system $h_{22}$ and $h_{33}$ do
not vanish and we get from the field equation (\ref{fe1}) with the
choice of indices
 $a,\  b=2$ and $a,\ b=3$ 
\begin{equation}{R_{22}\over{h_{22}}}={R_{33}\over{h_{33}}}=16\pi v^{-1}P\ .
\label{R33}\end{equation}
Therefore, the quantity $Pv^{-1}$, and thereby the left hand side of
(\ref{fe1}), depends merely on the
three-dimensional metric $h_{ab}$, just as in the linearly independent case of
\cite{ESI,R,RZS}. Following \cite{Geroch}, one may look for a transformation
which would generate from a known solution ($h_{ab},\ v,\ \omega$) a new solution
($h'_{ab},\ v',\ \omega'$) by leaving the the three-dimensional metric $h_{ab}$
intact. Since
$h_{ab}$ (and therefore $R_{ab}$) is supposed to be invariant against the
transformation, we get from (\ref{R33})
\begin{equation} P'v'^{-1}=Pv^{-1}\ , \label{pv}\end{equation}
where $P$ and $v$ are the initial pressure and norm, while $P'$ and
$v'$ denote the corresponding transformed quantities. This is one of the 
necessary conditions for the applicability of the
generalized transformation.

Next, in order to determine the exact form of the transformation
we recall the complex notation of \cite{Geroch}

\begin{equation}\tau=\omega+i v \ .\label{tau}\end{equation}
With this notation the field equations (\ref{fe1})-(\ref{fe3}) 
can be written as
\begin{equation}H_{ab}=-2(\tau-\bar\tau)^{-2}(D_{(a}\tau)(D_{b)}\bar\tau)\ ,
\label{Habcomplex}\end{equation}
\begin{equation} D_mD^m\tau=2(\tau-\bar\tau)^{-1}(D_m\tau )(D^m\tau )-i 8\pi
(\rho+3P) \ ,\label{Dtau} \end{equation}
where a bar denotes complex conjugation. Furthermore, in this complex notation the transformation we are looking
for is expected to have the form $h'_{ab}=h_{ab}$, $\tau '=\tau
'(\tau)$. It can be shown that,
just like in the vacuum case \cite{Geroch}, the
only solution to (\ref{Habcomplex}) and (\ref{Dtau}) of this form is
\begin{equation}\tau'={{a\tau+b}\over{c\tau+d}}\ ,
\label{tauprime}\end{equation}
where $a$, $b$, $c$ and $d$ are real constants which can be chosen,
without loss of generality, to satisfy $ad-bc=1$. From (\ref{tauprime})
 it follows immediately that
repeated applications of the transformation do not yield new 
(i.e. not gauge-related) solutions.
It was shown for the vacuum case 
 in \cite{Geroch} that by
factoring out with respect to gauge transformations from the three
parameters of an {\sl SL}(2,{I \hskip -4.5pt R})
group there remains only one independent
parameter.
Clearly, the same argument applies
 for the examined perfect fluid case as well.
One of the possible parameterizations is the choice $a=\cos\theta$, $b=\sin\theta$,
$c=-\sin\theta$ and $d=\cos\theta$, where now $\theta$ is the only independent parameter
of the transformation. Then, the real and imaginary parts
 of (\ref{tauprime}) give that
\begin{equation}v'={v\over{(\cos\theta-\omega \sin\theta)^2+
v^2\sin^2\theta}} ,\label{v}\end{equation}

\begin{equation}\omega'={{(\omega\cos\theta+\sin\theta)(-\omega\sin\theta+\cos\theta)-
v^2\sin\theta\cos\theta}\over{(\cos\theta-\omega\sin\theta)^2+v^2\sin^2\theta}} ,
\label{omega}\end{equation}
where $v'$ and $\omega'$ are 
the transformed norm and twist potential of the Killing field.

The transformation of the equation of state can be determined as follows.
The analogue of (\ref{Dtau}),
\begin{equation} D_mD^m\tau'=2(\tau'-\bar\tau')^{-1}(D_m\tau'
)(D^m\tau' )-8\pi(\rho'+3P')\ , \label{Dtauprime} \end{equation}
must hold, where $\rho'$ and $P'$
 are the transformed mass-density and pressure.
Calculating $D_m\tau'$ and $D_mD^m\tau'$ from (\ref{tauprime}) and substituting
into (\ref{Dtauprime}), using (\ref{tau}) we get a complex equation with the real part
\begin{equation}\rho+3P=(\rho'+3P')\bigl[ c^2\omega^2-c^2v^2+2cd\omega+d^2\bigr]\
,\label{re}\end{equation}
and imaginary part
\begin{equation}0=(\rho'+3P')cv(c\omega+d)\ . \label{im}\end{equation}

Obviously, the norm of the timelike Killing field, $v$, is non-zero,
as well as it is reasonable to suppose that the parameter of the transformation $\theta\not =0$ 
(i.e. $c \not =0$). Therefore, there remain
two possibilities so that (\ref{im}) be satisfied. First, we may assume that the 
equation of state of
the transformed perfect-fluid configuration is $\rho'+3P'=0$, which, in turn, gives
along with
 (\ref{re}) that $\rho+3P=0$. Unfortunately, this is
a highly unphysical equation of state. One would expect that a stationary
rigidly rotating perfect-fluid solution of Einstein's equations 
which gives a faithful representation of a
finite rigidly rotating body (e.g. a star) has to possess everywhere positive
mass-density and pressure in the
interior. Obviously, with $\rho+3P=0$ this  cannot  be the case.

There is, however, another possibility to have equation (\ref{im}) satisfied. 
Namely,
take $\omega=-{d\over c}={\cos\theta\over{\sin\theta}}=constant$, i.e. suppose that 
the initial spacetime is
static, allowing that $\rho'+3P' \not = 0$. The transformation of the equation of
state for this latter possibility can be determined as follows.

Without loss of generality, we may assume in the static case
that  $\omega=0$ which corresponds to
$\theta={\pi \over 2}$ (i.e. $c=-1$ and $d=0$).
 Then the transformation (\ref{v}), (\ref{omega}) reduces to
\begin{equation} v'={1\over v}\ ,\ \omega'=0\ , \label{vprime}\end{equation}
 yielding a new {\it static} solution.
This particular form of the transformation exactly reproduces
the Buchdahl transformation of perfect fluid spacetimes \cite{Buchdahl, Stewart} 
which has already been known for a long time.
The transformed mass-density, $\rho'$, and pressure, $P'$, can be determined as follows.
Substituting $\omega=0$, $c=-1$ and $d=0$ into (\ref{re}) we get
\begin{equation}\rho'+3P'=-{{\rho+3P}\over {v^2}}\ .
\label{rhopprime}\end{equation}
Moreover, from (\ref{pv}) and (\ref{vprime}) we get
\begin{equation}P'={P\over{v^2}}\ , \label{pprime}\end{equation}
which, along with (\ref{rhopprime}), gives
\begin{equation}\rho'=-{{\rho+6P}\over {v^2}}\ .\label{rhoprime}\end{equation}

The last two equations, exactly recover the corresponding equations
of \cite{Buchdahl, Stewart}. The transformed pressure
 and mass-density will both be positive only if the initial
pressure, $P$, is positive and the initial mass-density, $\rho$, satisfies
$\rho<-6P$ (compare with \cite{Stewart}).  In most cases these conditions can be easily
satisfied for an appropriate choice of parameters in the initial solution\footnote{
For example, with a suitable choice of parameters, the transformed pressure and
density can be made positive when one takes the interior (perfect-fluid) Schwarzschild solution
as the initial spacetime \cite{Exact} (see the discussion given by Stewart in
\cite{Stewart}). Further simple examples are the configurations obtainable from the metrics denoted
as Kuch71 II and R-R IV in \cite{DelgatyLake}.}.  However, for a physically meaningful
static spherical star model in addition to the positivity of the pressure and mass-density,
these quantities are expected to be monotonically decreasing functions of
the radial coordinate. It is also required that the solution be regular at the
origin\footnote{For the regularity conditions see e.g. \cite{DelgatyLake}.}.
Unfortunately, none of the metrics which we generated by applying
the above form of the Buchdahl
transformation to particular known solutions\footnote{These are the ones
denoted as Tolman I, II, V, VI,
Kuch2 VII, Kuch71 II,
 Whittaker, B-L, Heint IIIe, R-R I, III, IV, V, VI, Bayin
II and III in \cite{DelgatyLake}.} satisfy all criteria for
physical acceptability listed in \cite{DelgatyLake}.

\end{section}

\begin{section}{Herlt's transformation}

In this section we demonstrate that a technique given previously by Herlt in \cite{Herlt}
for generating stationary perfect-fluid
solutions from static ones is  a special case of the above described transformation. 
In particular, we show that the fundamental
equations of Herlt's transformation, equations (4.4a) and (4.4b) of
\cite{Herlt}, follow from the above considerations. Thereby, they  correspond
 to a special setting
of the examined transformation.

Since in \cite{Herlt} the initial spacetime is taken to be static, we write $\omega=0$
into (\ref{v}) and (\ref{omega}). This way we obtain

\begin{equation} v'={{(1+\tan^2\theta)v}\over{1+v^2\tan^2\theta}},
\label{cv}\end{equation}
\begin{equation} \omega'=
{{(1-v^2)\tan^2\theta}\over{1+v^2\tan^2\theta}}.\label{comega}\end{equation}

In the following we show that equations (4.4a-b) of \cite{Herlt}, i.e.
\begin{equation}D_a\bar U=
\sqrt{1+e^{-4U}\Bigl({{db}\over{dU}}\Bigr)^2}D_aU ,
\label{Ha}\end{equation}

\begin{equation}\bar pe^{-2\bar U}=pe^{-2U}, \label{Hb}
\end{equation}
 follow as a
special case from  (\ref{pv}), (\ref{cv}) and (\ref{comega})
of the present paper,
where our notation is related to the quantities $U$, $\bar U$, $p$ and $\bar p$ as 

\begin{equation} v=-e^{2\bar U}\ ,\  v'=-e^{2U}\ , \label{vU} \end{equation}
\begin{equation} \omega=0\ ,\ \omega '=2b\ ,\label{omb}\end{equation}
\begin{equation} P=\bar p\ , \ P'=p\ .\label{pprimep} \end{equation}

Now  equation (\ref{Hb}) (equation(4.4b) of \cite{Herlt}) follows immediately by
substituting  (\ref{vU})-(\ref{pprimep}) into 
(\ref{pv}).

To get  (\ref{Ha}) (equation (4.4a) of \cite{Herlt}), first we
substitute expressions (\ref{vU}) and (\ref{omb}) into 
(\ref{cv}), (\ref{comega}) and to the derivative of these
equations. Then, using the obtained system of equations, we
eliminate $e^{\bar U}$ and $\tan\theta$  from the equation
extracted by the substitution of (\ref{vU}) and (\ref{omb})
into (\ref{cv}). This way we obtain

\begin{equation} (d\bar
U)^2=\biggl[1+e^{-4U}\Bigl({{db}\over{dU}}\Bigr)^2\biggr] (dU)^2\
,\label{dUbar}\end{equation}
from which (\ref{Ha}) follows.
Note, however, that there is a sign ambiguity arising from (\ref{dUbar}),
 which is missing from (\ref{Ha})
(the original form of (4.4a) in \cite{Herlt}). Due to this, 
for instance in the particular case of static spacetimes,
$\omega'=2b=0$, the special transformation $dU=- d\bar U$ ($v'=1/v$ 
in our notation), which
recovers the Buchdahl transformation, was left out from
the analysis in \cite{Herlt}.

Let us, finally, mention that the equations used in \cite{Herlt}
(see equations (\ref{Ha}) and (\ref{Hb})) impose a strong
restriction on the possible form of the equation of state. Since Herlt's
transformation turned out to be a special case of the one examined throughout the previous
section, it is obvious that when the resulting
spacetime is non-static the relevant perfect-fluid configurations have to
possess the equation of state $\rho+3P=0$.

\end{section}

\begin{section}{Acknowledgments}
We wish to thank T. Dolinszky for carefully reading the
manuscript.
This research was supported in parts by the OTKA grants F14196
and T016246.

\end{section}

\eject

\end{document}